\documentclass[reprint,onecolumn,notitlepage,longbibliography]{revtex4-1}

\usepackage{amsmath,amssymb}    
\usepackage{graphicx}   
\usepackage{verbatim}   

\usepackage{color}      

\usepackage{hyperref}   

\usepackage{bm}
\usepackage{multirow}
\usepackage{upgreek} 
\usepackage{ulem}

\graphicspath{{./Figures/}}

\begin{document}
\title{Sharp-edged geometric obstacles in microfluidics promote deformability-based sorting of cells}
\author{Zunmin Zhang}
\author{Wei Chien}
\author{Ewan Henry}
\author{Dmitry A. Fedosov}\email{d.fedosov@fz-juelich.de}
\author{Gerhard Gompper}\email{g.gompper@fz-juelich.de}
\affiliation{Theoretical Soft Matter and Biophysics, Institute of Complex Systems and Institute for Advanced Simulation, 
Forschungszentrum J\"ulich, 52425 J\"ulich, Germany}
\date{\today}
	
\begin{abstract}
Sorting cells based on their intrinsic properties is a highly desirable objective, since changes in cell deformability are often associated 
with various stress conditions and diseases. Deterministic lateral displacement (DLD) devices offer high precision 
for rigid spherical particles, while their success in sorting deformable particles remains limited due to the complexity of cell 
traversal in DLDs. We employ mesoscopic hydrodynamics simulations and demonstrate prominent advantages of sharp-edged DLD obstacles for probing 
deformability properties of red blood cells (RBCs). By consecutive sharpening of the pillar shape from circular to diamond 
to triangular geometry, a pronounced cell bending around an edge is achieved, serving as a deformability sensor. Bending around
the edge is the primary mechanism, which governs the traversal of RBCs through such DLD device. This strategy requires an appropriate 
degree of cell bending by fluid stresses, which can be controlled by the flow rate, and exhibits good sensitivity to moderate changes
in cell deformability. We expect that similar mechanisms should 
be applicable for the development of novel DLD devices that target intrinsic properties of many other cells. 
\end{abstract}

\maketitle

\section{Introduction}

Deterministic lateral displacement (DLD) is a powerful microfluidic technique capable of continuous size-dependent particle sorting with 
exceptional resolution \cite{Huang_DLD_2004,Inglis_CPS_2006,McGrath_DLD_2014}. Due to the extreme precision capabilities, DLDs promise to have 
great potential for biological and clinical applications. DLD devices have been employed to sort out circulating tumor cells from blood 
\cite{Loutherback_SCC_2012,Liu_RIC_2013,Karabacak_MIC_2014}, to process blood cells \cite{davi06,Beech_SCS_2012,Zeming_RSP_2013,Ranjan_DLD_2014}, 
and to separate bacteria \cite{Ranjan_DLD_2014,Beech_SPB_2018} and parasites \cite{holm11,Holm_SMS_2016} from human blood. However, the auspicious 
precision is directly realizable only for rigid spherical particles, while biological particles (or cells) are generally deformable and non-spherical \cite{Zeming_RSP_2013,Henry_SCP_2016}. Therefore, 
biological cells may attain different flow-induced orientation and deform in response to local flow conditions, making the prediction of their traversal 
through a DLD device difficult. In fact, the success in several cases above can mainly be attributed to significant size differences 
between targeted bioparticles and other background cells. Nevertheless, in numerous  biological and clinical applications sorting and detection based on 
intrinsic mechanical properties of cells is highly coveted, since cell's deformability is recognized as an important biomarker for the state of a cell \cite{Bao_CMM_2003,DiCarlo_MBC_2012}.
For instance, a gradual stiffening of red blood cells (RBCs) is an important indicator for the onset and progression of diseases such as diabetes, 
sickle cell anemia, and malaria \cite{Kaul_Fabry_1983,Miller_PBM_2002,Cranston_PFM_1984}. This motivates large scientific efforts to facilitate 
precise sorting in DLDs using intrinsic cell properties such as shape and deformability. 

DLD devices usually utilize a row-shifted post (or pillar) array, which divides the laminar flow into several streams with equal volumetric flow rates \cite{Inglis_CPS_2006}. The thickness of the first 
stream adjacent to a pillar defines a critical radius $R_{\rm c}$ such that rigid spherical particles with a larger or smaller size than $R_{\rm c}$ 
traverse in a displacement or zigzag mode, respectively \cite{Inglis_CPS_2006,Kulrattanarak_MMD_2011,Kim_BFS_2017}. Recently, an intermediate traversal mode 
called 'mixed' mode has also been demonstrated experimentally \cite{Kulrattanarak_AMM_2011,Zeming_DLD_2016}, theoretically \cite{Kulrattanarak_MMD_2011,Kim_BFS_2017},  
and in simulations \cite{Henry_SCP_2016,Kulrattanarak_AMM_2011,Zhang_BRD_2015}. Biological cells can attain not only all these modes, but also change their effective critical size 
depending on the flow conditions \cite{Beech_SCS_2012,Henry_SCP_2016,Krueger_DBS_2014,Quek_DLD_2011}. For example, RBC orientation can be sterically 
controlled in a shallow DLD device whose depth is smaller than the large diameter of the discocytic shape \cite{Beech_SCS_2012,holm11,Holmes_SBC_2014}, and cell deformation 
can be mediated by different fluid stresses at various flow rates \cite{Beech_SCS_2012,Krueger_DBS_2014,Holmes_SBC_2014}. However, shallow devices have a low throughput and often 
suffer from cell sticking at the walls. 
Recently, an asymmetric deep DLD device (i.e. its depth is larger than the largest particle dimension) with a reduced inter-row gap size \cite{Zeming_DLD_2016} has been shown to provide a better sorting of 
RBCs, wherein the underlying mechanism resembles cross-flow filtration \cite{Ripperger_CFM_2002}. Different post geometries 
have also been used to explore their possible advantages. For instance, I-shaped and L-shaped pillars were suggested as good candidates for sorting non-spherical 
particles, as they can induce a rotational motion \cite{Zeming_RSP_2013,Ranjan_DLD_2014}. In our previous work \cite{Zhang_BRD_2015}, we have demonstrated 
that RBCs exhibit different deformation in arrays with various pillar geometries, since pillar shape influences the flow field in a DLD. Airfoil-like and diamond 
post shapes have been suggested to be advantageous for handling soft bioparticles, as they minimize device clogging \cite{AlFandi_DLD_2011}. 
Nevertheless, there is no comprehensive understanding what mechanisms and post structures should be employed to establish an efficient sorting scheme
which would target cell deformability as a biomarker.  
   
We employ mesoscale simulations to identify robust mechanisms and advantages of different post geometries for deformability-based sorting of RBCs. 
Our main result is that sharp-edged structures in deep DLDs are able to significantly enhance sensitivity for moderate differences in RBC deformability. 
Conventional deep DLD arrays based on circular pillars have a poor performance for deformability-based sorting, because the rounded post shapes do not 
generate distinct enough deformation of cells with different elastic properties. In contrast, diamond and triangular structures with sharp edges induce strong RBC bending around a post, which is the main mechanism for excellent deformability sensitivity, and thus, serves as a sensor for cell's deformability. 
In addition, sharp edges result in a bended first stream, whose length along the flow direction is significantly reduced. The relation between the 
length of the first stream in the flow direction and a deformability-dependent effective length of RBCs significantly influences the traversal of RBCs
through the device. This constitutes a novel sorting mechanism, which is important for the separation of  deformable and anisotropic particles. 
It is important to emphasize that optimal sensitivity of different sharp-edged structures is achieved only for an appropriate correspondence between cell deformation 
and fluid stresses applied by the flow, so that the flow rate has to be carefully selected. 
Our results provide general design principles for DLD devices, and should thus be useful for 
the construction of optimized microfluidic devices, which can precisely target intrinsic cell properties.

\section{Simulation methods and models}

\subsection{Post geometry and arrangement in DLDs}

\begin{figure}[!t]
\centering
\includegraphics[width=0.6\linewidth]{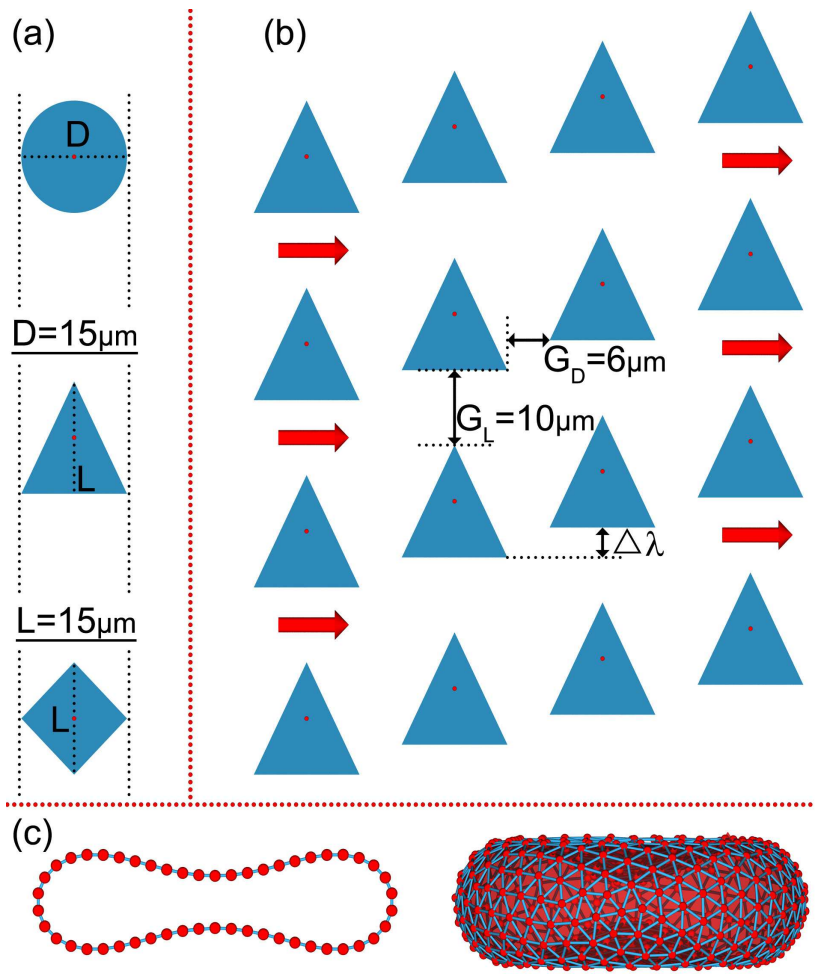}
\caption{{\bf Schematic of DLD devices and RBC models.} (a) Pillar shapes including circle, triangle, and diamond geometries
characterized by the pillar sizes $L = D = 15$ $\upmu\rm m$. (b) Asymmetrical DLD array with triangular pillars characterized 
by the asymmetric gap sizes (lateral gap of $G_{\rm L} = 10$ $\upmu\rm m$ and downstream gap of $G_{\rm D} = 6$ $\upmu\rm m$), 
the post center-to-center distances $\lambda_L=L + G_L$ and $\lambda_D=D + G_D$, and a row shift $\Delta\lambda$. 
Red arrows indicate the average flow direction along the x-axis of the simulated box, while the row 
	shift is made along the y-axis. Note that in 3D simulations, the device thickness is represented by 
	the length of simulated systems along the z-axis. (c) 2D bead-spring model and 3D triangular mesh model of a RBC.}
\label{fig1}
\end{figure}

To demonstrate how sharp-edged obstacles in deep DLD devices can dramatically enhance 
deformability-based sorting of RBCs, we study microfluidic DLD arrays with
different post geometries (see Fig.~\ref{fig1}). Investigated post shapes include circular, diamond, and triangular obstacles with two characteristic 
lengths $L=D=15$ $\upmu\rm m$, as depicted in Fig.~\ref{fig1}(a). The corresponding array gaps are $G_{\rm L}$ and 
$G_{\rm D}$ with a row shift  $\Delta \lambda$ and the post center-to-center distances are $\lambda_L=L + G_L$ 
and $\lambda_D=D + G_D$, as shown in Fig.~\ref{fig1}(b). 
The DLD array we study in simulations is asymmetric ($G_{L}:G_{D}=10:6$), which 
leads to a reduced critical size in comparison to a conventional symmetric design with $G_{D}=G_{L}$ 
\cite{Zeming_DLD_2016}.   

We employ simulations of two-dimensional (2D) and three-dimensional (3D) systems to study 
RBC behavior in DLD devices for a wide range of parameters. 
The 2D model is employed for its numerical efficiency; however, we will show 
that it qualitatively captures RBC behavior in comparison with the 3D model. 

\subsection{RBC model}

In the simulations, a RBC in 2D is represented by a bead-spring chain, incorporating 
bending rigidity ($\kappa$) and length and area conservation, while the membrane in 3D 
is modeled by a triangular mesh, 
which represents shear elasticity, bending rigidity, and area and volume conservation, see Fig.~\ref{fig1}(c).     

In 2D, RBCs are modeled as closed bead-spring chains with $N_{p}=50$ beads 
connected by $N_{s}=N_{p}$ springs \cite{Zhang_BRD_2015,Fedosov_WBC_2012}. 
The spring potential is given by
\begin{equation}\label{mm2}
U_{sp} = \sum_{j=1}^{N_{s}} \left[\frac{k_{B}Tl_{m}(3x_{j}^{2}-2x_{j}^{3})}{4l_{p}(1-x_{j})} + \frac{k_{p}}{l_{j}}\right],
\end{equation}
where $l_{j}$ is the length of the spring $j$, $l_{m}$is the maximum spring extension, $x_{j}=l_{j}/l_{m}\in (0, 1)$, $l_{p}$ is the persistence length, and $k_{p}$ is the spring constant. Note that 
a balance between the two force terms in Eq.~(\ref{mm2}) results in a non-zero equilibrium spring length $l_{0}$, with $l_{m}/l_{0}=2.2$. 
Furthermore, to induce a 2D biconcave cell shape in equilibrium, an additional membrane bending potential $U_{bend}$ and area constraint 
potential $U_{area}$ are incorporated as 
\begin{equation}\label{mm3}
\begin{split}
U_{bend} &= \sum_{j=1}^{N_{p}} k_{b}[1-\cos(\theta_{j})], \\
U_{area} &= k_{a} \frac{(A-A_{0})^{2}}{2},
\end{split}
\end{equation}
where $k_{b}$ and $k_{a}$ are the bending coefficient and the area-constraint coefficient, respectively. 
$\theta_{j}$ is the instantaneous angle between two adjacent springs having a common vertex 
$j$, $A$ is the instantaneous area, and $A_{0}$ is the desired cell area. In 2D, $k_b$ is related to the 
macroscopic bending rigidity $\kappa$ as $\kappa=k_b l_0$.    

Typical values for a healthy RBC are the effective cell diameter 
$D_{r}^{2D} = 6.1\upmu m$ and the bending rigidity $\kappa$ of about $50-70k_{B}T$ for the physiological 
temperature $T=37^o$C.  
In 2D, we use $D_{r}^{2D} = L_{r} /\pi$, with the cell contour length $L_{r}$. 
All 2D simulation parameters are collected in Table \ref{table:method}. The RBC bending rigidity is 
characterized by 
a non-dimensional rigidity factor $K_{2D}^*=\kappa/\kappa_{0} \in \{1, 2, 4, 10, 20, 40\}$, where 
$\kappa_0/(k_B T l_0) = 50$ corresponds to the typical bending rigidity of a healthy RBC. 

In 3D, the RBC membrane is modeled as a triangulated network of springs 
\cite{gompper2004triangulated,Noguchi_STV_2005,Fedosov_RBC_2010,Fedosov_SCG_2010,Fedosov_MBF_2014}. It is constructed
from a collection of $N_{v}=1000$ particles linked by $N_{s}=3(N_{v}-2)$ springs with the bonding potential 
of Eq.~(\ref{mm2}). The membrane bending rigidity in absence 
of spontaneous curvature is described by a bending energy  
\begin{equation}
U_{bend} = \sum_{j=1}^{N_{s}} k_{b}[1-\cos(\phi_{j})],
\end{equation}
where $\phi_j$ is the angle between two normal vectors of triangular 
faces adjacent to the spring $j$. In addition, to represent area-incompressibility of the lipid bilayer and incompressibility of the inner cytosol, the area and volume constraints are introduced by the potentials
\begin{equation}\label{mm5}
\begin{split}
U_{area} & = k_{a}\frac{(A-A_{r})^2}{2A_{r}} + \sum_{j \in 1...N_{t}} k_{d}\frac{(A_{j}-A_{j}^{0})^2}{2A_{j}^{0}}, \\
U_{vol} &= k_{v}\frac{(V-V_{r})^2}{2V_{r}},
\end{split}
\end{equation} 
where $k_{a}$, $k_{d}$, and $k_{v}$ are the global area, local area, and volume constraint coefficients, 
respectively. $A$ and $V$ are the instantaneous area and volume of RBCs 
with the desired values of $A_{r}$ and $V_{r}$, while $A_{j}$ and $A_{j}^{0}$ are the instantaneous and desired areas of the $j$-th triangle in the network, respectively. 
Note that the desired values of those parameters are set according to the initial triangulation.

The parameters of the triangulated-network model have been related to the macroscopic membrane properties 
\cite{Seung_DFM_1988,Gompper_1996,Fedosov_RBC_2010,Fedosov_SCG_2010}. 
For example, the membrane shear modulus is given by
\begin{equation}\label{mm6}
\mu_{r} = \frac{\sqrt{3}k_{B}T}{4pl_{m}x_{0}} (\frac{x_{0}}{2(1-x_{0})^3}-\frac{1}{4(1-x_{0})^2}+\frac{1}{4}) + \frac{2\sqrt{3}k_{p}}{4l_{0}^{3}},
\end{equation}
where $x_{0}=l_{m}/l_{0}=2.2$ for all springs. The bending constant $k_{b}$ can be expressed in terms of the macroscopic bending 
rigidity $\kappa$ of the Helfrich model \cite{Helfrich_EPB_1973} as $\kappa=\sqrt{3}k_{b}/2$. Accordingly, mechanical properties of 3D RBCs can be characterized by both 
the shear modulus $\mu_{r}$ and bending rigidity $\kappa$ with a F\"oppl-von K\'arm\'an number $\Gamma =\mu_{r} D_{r}^2 /\kappa$. In this study, $\mu_{r}$ is fixed and 
$\kappa$ is varied and characterized by $K_{3D}^{*} = \kappa/ \kappa_{0}$, where $\kappa_{0} = 70 k_B T$ corresponds to a healthy RBC with $\Gamma \approx 680$. 
An effective RBC diameter is defined as $D_{r} = \sqrt{A_{r}/\pi}$ with $A_{r}=133.5 \times 10^{-12} m^2$, implying that $D_{r} = 6.5\upmu m$. RBC volume is set to 
$V_{r}=94 \times 10^{-18} m^3$. Table \ref{table:method2} summarizes important parameters used in 3D simulations.

\subsection{Mesoscale hydrodynamics approach}

We employ the dissipative particle dynamics (DPD) method \cite{Hoogerbrugge_SMH_1992,Espanol_SMO_1995} 
in 2D to qualitatively explore physical mechanisms for efficient deformability-based separation. 
The smoothed dissipative particle dynamics (SDPD) approach \cite{Espanol_SDPD_2003,Mueller_SDPD_2015} 
is applied in 3D for a detailed quantitative exploration of specific flow conditions and 
post shapes.

In the DPD method \cite{Hoogerbrugge_SMH_1992,Espanol_SMO_1995}, each individual particle represents a group of molecules. 
The inter-particle force (${\bm{F}_{ij}}$) acting on particle $i$ by particle $j$ is a sum of three pairwise forces, 
conservative (${\bm{F}}_{ij}^{C}$), dissipative (${\bm{F}}_{ij}^{D}$), and random (${\bm{F}}_{ij}^{R}$) within a selected cutoff radius $r_{c}$.
\begin{equation}\label{mm1}
\begin{split}
{\bm{F}}_{ij}^{C}&= a_{ij}(1-r_{ij} / r_{c})\hat{{\bm{r}}}_{ij},\\
{\bm{F}}_{ij}^{D}&= -\gamma_{ij} \omega^{D} (r_{ij}) ({\bm{v}}_{ij} \cdot \hat{{\bm{r}}}_{ij}) \hat{{\bm{r}}}_{ij},\\	
{\bm{F}}_{ij}^{R}&= \sigma_{ij} \omega^{R} (r_{ij}) \xi_{ij} dt^{-1/2} \hat{{\bm{r}}}_{ij},
\end{split}
\end{equation}
where ${\bm{r}}_{ij} = {\bm{r}}_{i} - {\bm{r}}_{j}$, $r_{ij} = |{\bm{r}}_{ij}|$, $\hat{{\bm{r}}}_{ij} = {\bm{r}}_{ij} / |{\bm{r}}_{ij}|$ and ${\bm{v}}_{ij} = {\bm{v}}_{i} - {\bm{v}}_{j}$. The coefficients $a_{ij}$, 
$\gamma_{ij}$, and $\sigma_{ij}$ determine the strength of the corresponding forces, respectively. $\omega^{D} (r_{ij})$ and $\omega^{R} (r_{ij})$ are weight functions, and $\xi_{ij}$ is a random 
number generated from a Gaussian distribution with zero mean and unit variance. The dissipative and random forces must satisfy the fluctuation-dissipation theorem given by the conditions 
$\omega^{D}(r_{ij}) = [\omega^{R}(r_{ij})]^{2}$ and $\sigma^{2} = 2\gamma k_{B}T$ \cite{Espanol_SMO_1995}, in order to maintain the system temperature ($T$) and generate a correct equilibrium 
Gibbs-Boltzmann distribution. $\omega^{R} (r_{ij})=(1-r_{ij}/r_{c})^k$ with $k=0.15$ is employed in this study. 
The DPD parameters for 2D simulations are given in Table \ref{table:method}.

\begin{table}
	\begin{tabular}{c c c c c}
	\hline
	\multirow{2}{*}{DPD} & $n r_c^2$ & $ar_c/k_BT$ & $\gamma r_c/\sqrt{mk_BT}$ & $\eta r_c/\sqrt{mk_BT}$\\
	&$11.25$ & $60$ & $30$ & $216.7$\\
	\hline
	\multirow{2}{*}{RBC} & $A_0/D_{r}^2 $ & $\kappa/(k_{B}T l_0)$ & $YD_{r}/k_BT$ & $k_{a}D_{r}^{2}/k_BT$\\
	& $0.36$ & $50 \cdot K_{2D}^{*}$ & $9000 \cdot K_{2D}^{*}$ & $37430 \cdot K_{2D}^{*}$\\
	\hline
	\end{tabular}
\caption{Fluid and cell parameters used in 2D simulations. $n$ is the fluid number density and $Y=(-\partial^2 U_{sp}/\partial l^2)|_{l_0}$ is the stretching modulus with $l_0=0.26r_c$ being 
equilibrium spring length.  
Note that mass, length, and energy are measured in units of the fluid particle mass $m = 1$, the cutoff radius $r_c =1.5$, and $k_{B}T = 1.0$, respectively.}
\label{table:method}
\end{table}

In the SDPD method \cite{Espanol_SDPD_2003,Mueller_SDPD_2015}, each individual particle represents a small volume of fluid. SDPD is often considered as an 
improved version of DPD with a direct connection to the Navier-Stokes equation and consistent thermal fluctuations \cite{Quesada_CSF_2009}. The three pairwise 
forces on particle $i$ in SDPD are given by 
\begin{equation}\label{mm4}
\begin{split}
{\bm{F}}_{i}^{C}&= \displaystyle\sum_{j}(\frac{p_{i}}{\rho_{i}^{2}}+\frac{p_{j}}{\rho_{j}^{2}})\omega_{ij}\hat{{\bm{r}}}_{ij},\\
{\bm{F}}_{i}^{D}&= -\displaystyle\sum_{j}\gamma_{ij}({\bm{v}}_{ij} +({\bm{v}}_{ij}\cdot\hat{{\bm{r}}}_{ij})\hat{{\bm{r}}}_{ij}),\\
{\bm{F}}_{i}^{R}&= \displaystyle\sum_{j}\sigma_{ij} (d\overline{\mathbf{W}}_{ij}^{S}+\frac{1}{3}tr[d\mathbf{W}_{ij}])\cdot\hat{{\bm{r}}}_{ij},
\end{split}
\end{equation}
where $tr[d\mathbf{W}_{ij}]$ and $d\overline{\mathbf{W}}_{ij}^{S}$ are the trace of a random matrix of independent Wiener increments $d\mathbf{W}_{ij}$ and its traceless 
symmetric part, respectively. $p_{i}$ and $p_{j}$ are local particle pressures given by the equation of state $p=p_{0}(\rho/\rho_{0})^{7}-b$, where $p$, $\rho_{0}$ and $b$ 
are selected parameters. $\gamma_{ij}$ and $\sigma_{ij}$ are the coefficients of dissipative and random forces respectively, such that 
$\gamma_{ij}=\frac{5\eta_{0}}{3}\frac{\omega_{ij}}{\rho_{i}\rho_{j}}$ and $\sigma_{ij}= 2\sqrt{k_{B} T \gamma_{ij}}$ with $\eta_{0}$ being the desired dynamic viscosity. 

\begin{table}[h]
\centering
\begin{tabular}{c c c c c c }
	\hline
	\multirow{2}{*}{SDPD} & $nr_c^3$ & $p_{0}r_c^3/k_B T$ & $b r_c^3/k_B T$  & $\eta r_{c}^2/\sqrt{mk_BT}$\\
	&$20.25$ & $25313$ & $24975$  & $503.1$  \\
	\hline
	\multirow{2}{*}{RBC} & $N_{v}$ & $l_{m} /l_{0}$ & $\kappa/k_B T$ & $\mu_{r}D_r^2/k_B T$  \\
	& $1000$ & $2.2$ & $70\cdot K_{3D}^{*}$ & $47620$ \\
	\hline
	\end{tabular}
\caption{Fluid and cell parameters used in 3D simulations. In all simulations, we set $m = 1$, $r_c =1.5$, and $k_{B}T = 0.2$.}
\label{table:method2}
\end{table}

\subsection{Simulation characteristics}
\label{sec:sim}

A simulated system consists of a single suspended RBC, a single pillar, and many fluid particles within the 
computational domain. Post walls and 
the ceiling and floor walls in 3D are modeled by a layer of frozen particles with a thickness of $r_{c}$ and same equilibrium structure as that of 
the suspending fluid. In addition, to prevent wall penetration, bounce-back reflections are applied to particles at fluid-solid interfaces. An infinite DLD array is represented 
by applying periodic boundary conditions (BCs) both in the flow (x) and lateral (y) directions. To model a shift $\Delta \lambda$ in the lateral 
direction, a shift in the $y$  direction is applied for each boundary-crossing event in the $x$ direction. 

The flow in the $x$ direction is driven by a constant force $f$ applied to each fluid particle, which is equivalent to a constant 
pressure gradient $\Delta P/ \lambda_D = fn$, where $\Delta P$ is the pressure drop over a single post column and $n$ is 
the number density of fluid particles in simulations. To characterize the flow strength, we define a Capillary number 
as $Ca = \bar{\dot{\gamma}}\eta D_{r}^3/ \kappa$. Here, the average shear rate is $\bar{\dot{\gamma}}=Q/G_L^2$ in 2D 
and $\bar{\dot{\gamma}}=Q/(G_L^2 H_{DLD})$ in 3D, where $Q$ is the flow rate within a single post
section and $H_{DLD}$ is the height of a DLD device in 3D. $\eta$ is fluid's dynamic viscosity.
In 3D simulations, another capillary number based on the shear modulus of RBC membrane can be defined 
as $Ca_{\mu}=\bar{\dot{\gamma}}\eta D_{r}/ \mu_r$,  which is related to the employed Ca number based on the bending rigidity through 
the F\"{o}ppl-von K\'{a}rm\'{a}n number $\Gamma \approx 680$. Which of these capillary numbers is most relevant depends on 
the type of deformation. For bending of RBCs around sharp edges considered here, it is $Ca$, whereas for RBC buckling in shear flow
at high shear rates, it is $Ca_{\mu}$ \cite{Mauer_FIT_2018}. 
It is important to note that an additional adaptive force is applied in $y$ direction to ensure no net flow 
in the lateral direction \cite{Henry_SCP_2016,Zhang_BRD_2015}. Both 2D and 3D simulations are performed with a viscosity contrast $C=\eta_{i}/\eta=1$ between the internal ($\eta_i$) and 
external ($\eta$) fluid viscosities. 

To analyze cell deformation, we employ RBC acircularity $\delta_{2D}$ in 2D \cite{Zhang_BRD_2015} and asphericity $\delta_{3D}$ in 3D \cite{Fedosov_PBV_2011},
which are defined as
\begin{equation}
\delta_{2D} = \frac{(\lambda_{1}-\lambda_{2})^2}{(\lambda_{1}+\lambda_{2})^2} \ ,
\end{equation}
\begin{equation}
\delta_{3D} = \frac{[(\lambda_{1}-\lambda_{2})^2 + (\lambda_{2}-\lambda_{3})^2 + (\lambda_{3}-\lambda_{1})^2]}{2(\lambda_{1}+\lambda_{2} + \lambda_{3})^2} \ ,
\end{equation}
where $\lambda_{1} \geq \lambda_{2} \geq \lambda_{3}$ are the square roots of the eigenvalues of the squared radius-of-gyration tensor. Acircularity (or asphericity) characterizes 
the departure of cell shape from a circular (or spherical) geometry. Thus, the value of $\delta$ is equal to zero for a perfect circular or spherical shape, and approaches $1.0$ for 
a strongly elongated object. In equilibrium, $\delta_{2D} \approx 0.29$ and $\delta_{3D} \approx 0.13$ correspond to the 2D and 3D 
biconcave shapes of a RBC, respectively.

\section{Results}

\subsection{RBC behavior in circular post arrays}

\begin{figure}[!t]
\centering
\includegraphics[width=\linewidth]{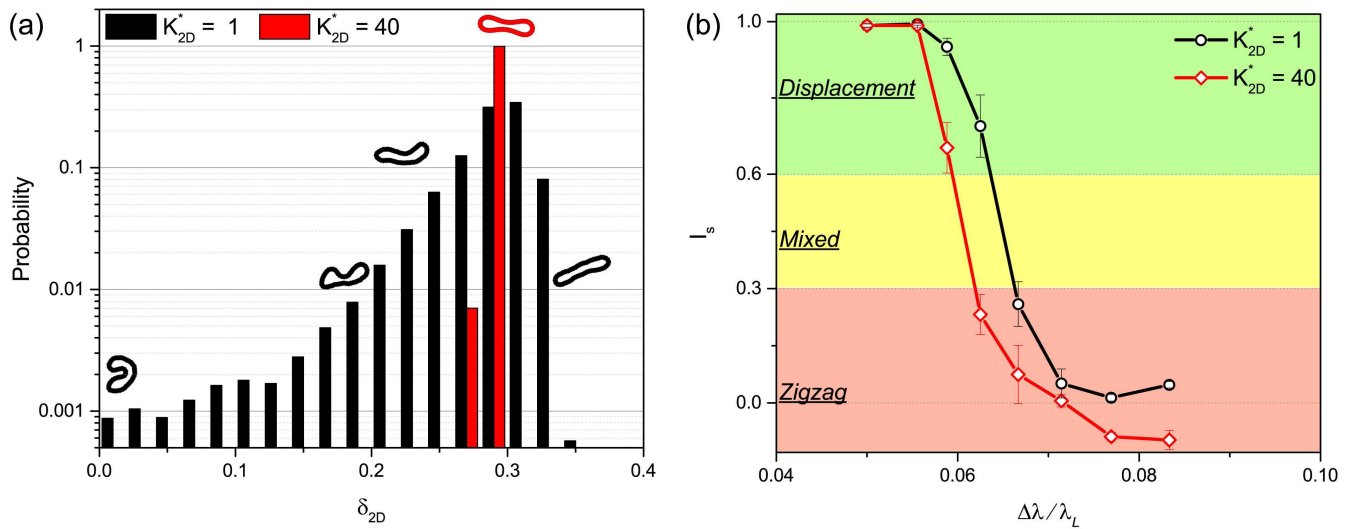}
\caption{{\bf Deformation and displacement of RBCs in circular post array.} (a) Acircularity distributions of soft ($K^*_{\rm 2D}=\kappa/ \kappa_0  = 1$) 
and stiff ($K^*_{\rm 2D} = 40$) RBCs in a circular post array with $\Delta \lambda / \lambda_L = 0.0625$. The acircularity is defined as 
$\delta_{2D} = (\lambda_{1}-\lambda_{2})^2/(\lambda_{1}+\lambda_{2})^2$, where $\lambda_{1} \geq \lambda_{2}$ are square roots of the eigenvalues 
of the squared radius-of-gyration tensor.  The plotted shapes of RBCs 
represent typical cell configurations from simulated trajectories. (b) Separation index $I_{s}$ for a RBC as a function of the row shift $\Delta \lambda$. 
$I_s$ is the ratio of average lateral displacement of a RBC per post and the row shift $\Delta \lambda$. Error bars correspond to 
standard deviations. Here, $Ca = 34.1/ K^*_{\rm 2D}$.} 
\label{fig2}
\end{figure}

We start with a conventional circular post geometry and investigate its performance for deformability-based sorting of RBCs. Fig.\ref{fig2}(a) presents 
acircularity distributions, which characterize the deviation of cell shape from a circular configuration for the two very different RBC bending rigidities 
($K^*_{\rm 2D}=\kappa/ \kappa_0  = 1$ and $40$). The softer RBC with bending rigidity $\kappa_0$ (i.e. $K^*_{\rm 2D}=1$)  approximates normal healthy conditions (see Methods) and shows 
considerable deformations in DLD, documented by a wide acircularity distribution in Fig.\ref{fig2}(a). The stiffer cell exhibits a rather narrow acircularity 
distribution, which is centered around an acircularity value $\delta_{\rm 2D} \approx 0.29$ of the RBC biconcave shape in equilibrium, indicating that a RBC 
with $K^*_{\rm 2D}  = 40$ can be considered nearly rigid under these flow conditions with $Ca = 34.1/ K^*_{\rm 2D}$. Such strong differences in the deformation of these two cells should 
in general strongly affect their traversal through the DLD device and eventually translate into their tangible separation.            

To quantitatively characterize the traversal of RBCs through a device, a dimensionless parameter called "separation index" $I_{\rm s}$, defined as the ratio 
of average lateral displacement of a cell per post and the row shift $\Delta \lambda$, is introduced similarly to Ref.~\cite{Zhang_BRD_2015}. An ideal displacement 
mode is characterized by $I_{\rm s} =1$ such that cells are laterally displaced at each post. An ideal zigzag mode is generally referred to as $I_{\rm s} \approx 0$ without 
a significant net lateral displacement of cells in the device. In addition, a "mixed mode" with irregular alternation of zigzag and displacement modes can be 
also present, for which $I_{\rm s} \in [0.3, 0.6]$ \cite{Zhang_BRD_2015}, and be used to localize the displacement-to-zigzag transition. Fig.~\ref{fig2}(b) shows 
the separation index for soft and nearly rigid RBCs as a function of $\Delta \lambda$.  Despite the very strong differences in overall deformation of these 
two RBCs illustrated in Fig.\ref{fig2}(a), $I_{\rm s}$ displays only a slight shift toward a lower value of $\Delta \lambda$ with increasing cell rigidity. This results in a very 
narrow window of $\Delta \lambda$ values ($\sim 100$ nm), where deformability-based sorting should theoretically be possible. However, in practice the use of 
this narrow separation window would likely lead to a poor performance or inability to efficiently carry out deformability-based cell sorting. Therefore, strong differences 
in particle deformation within a DLD device do not guarantee good or even moderate sorting efficiency.  

\begin{figure}[!t]
\centering
\includegraphics[width=0.6\linewidth]{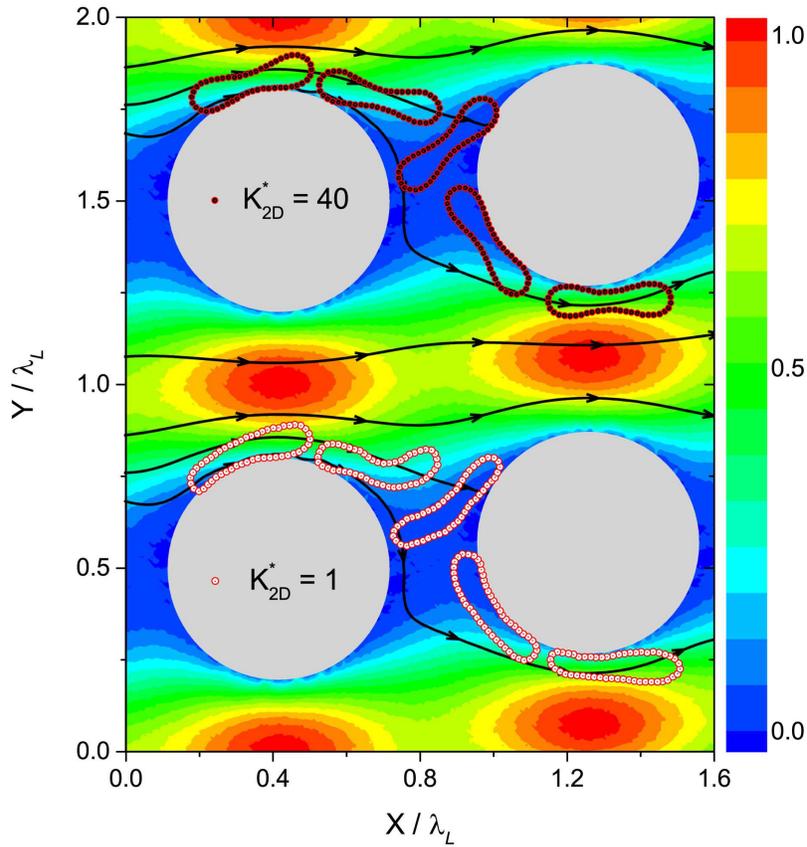}
\caption{{\bf Motion and deformation of soft ($K^*_{\rm 2D}  = 1$, bottom) and  nearly rigid ($K^*_{\rm 2D}  = 40$, top) RBCs during a zigzagging 
event.} 
The snapshots are extracted from simulated flow trajectories of RBCs in a circular post array with $\Delta \lambda  / \lambda_L = 0.0625$. 
In the background, corresponding fluid flow profiles of the velocity in flow direction ($x$ axis) are presented with several typical streamlines. 
The velocity field is normalized by a maximum value which is same for both cases. $Ca = 34.1/ K^*_{\rm 2D}$.}
\label{fig3}
\end{figure}

To identify cell deformation which primarily governs the traversal of RBCs through a DLD device, we take a closer look at RBC trajectories (Fig.~\ref{fig3}). Such visual analysis reveals 
that RBC deformation near the top of a pillar determines cell's trajectory, as it can result in a further displacement motion or initiate a zigzaging event. This is also 
consistent with previous studies \cite{Krueger_DBS_2014,Beech_SCS_2012,Henry_SCP_2016}, which demonstrated that strong enough deformation near a post 
can effectively reduce particle size and alter its traversal through a DLD device. Note that the argument about effective reduction of the particle size due to deformability
is not strictly applicable here, since the soft cell displays a slightly larger effective size than the stiff RBC in Fig.~\ref{fig2}(b). Figure \ref{fig3} compares typical motion of soft 
and stiff RBCs as they experience a zigzagging event. Despite some differences in cell deformation, the motion of both RBCs is quite similar. In fact, significant deformations of 
soft RBCs represented by the wide acircularity distribution in Fig.~\ref{fig2}(a) occur mainly away from the posts, and do not significantly contribute to cell's traversal through 
the DLD device. Interestingly, slight deformations of soft RBCs near a post in Fig.~\ref{fig3} aid the RBC to escape from the first stream, leading to a larger effective size and $I_{s}$ value 
in Fig.\ref{fig2}(b) in comparison to the stiff cell. However, it is apparent that the rounded geometry of circular posts and the resulting flow field do not induce 
strong enough differences in RBC deformation near a post and thus, result in a poor deformability-based separation.

\begin{figure}[!t]
\centering
\includegraphics[width=0.6\linewidth]{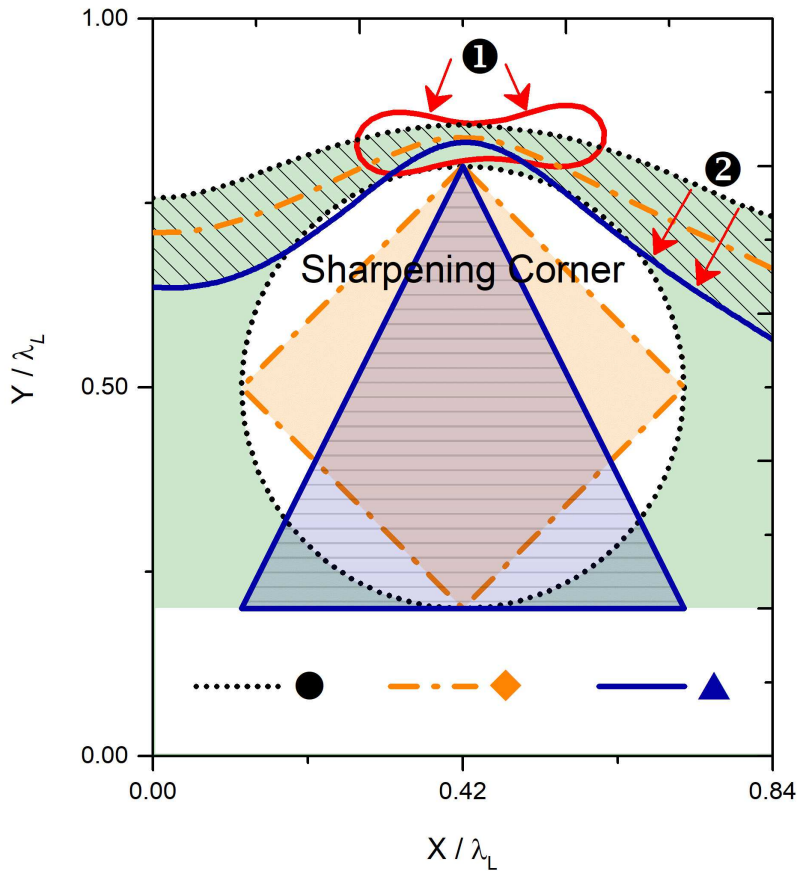}
\caption{{\bf Schematic of different post shapes.} Different pillar shapes illustrate a gradual edge sharpening from circular to diamond to triangular geometry. 
Furthermore, various post shapes lead to differences in bending of RBCs (1) and stall lines (2) corresponding to the first stream of the flow in the DLD device.}
\label{fig4}
\end{figure}

\subsection{Sorting with sharp-edged post structures}

The results for circular posts indicate that a considerable deformation of cells near the top of post structures likely has a strong correlation with the 
sensitivity of a DLD device for deformability-based sorting. A stronger RBC deformation can be initiated by sharper edges at the post top, which is illustrated 
by different post geometries in Fig.~\ref{fig4}, including circular, diamond, and triangular structures. Sharp edges in post geometry result in two crucial changes 
for RBC behavior in DLD. First, such structures are able to induce a considerable RBC bending around a post due to steric cell-post interactions and corresponding 
flow stresses within a flow field with bent streamlines, since a RBC generally deforms along a post edge \cite{Zeming_DLD_2016}. Second, a sharp edge leads to 
a reduced thickness of the first stream along this edge, as illustrated in Fig.~\ref{fig4}. For rigid spherical particles, this reduces the critical size for separation
\cite{Loutherback_IPD_2010,Zhang_BRD_2015}. Furthermore, Fig.~\ref{fig4} displays that a length of the bent first 
stream adjacent to the edge is significantly reduced for sharp edges, which is important in comparison to a characteristic RBC length. Note that the 
length of the  bent first stream is also reduced due to the asymmetric post positioning with a smaller gap $G_D$ between consecutive rows. As it will become evident further on, 
for the separation of non-spherical particles, the stream length instead of its thickness may play a decisive role for particle traversal through a DLD device. 

\begin{figure}[!t]
\centering
\includegraphics[width=0.6\linewidth]{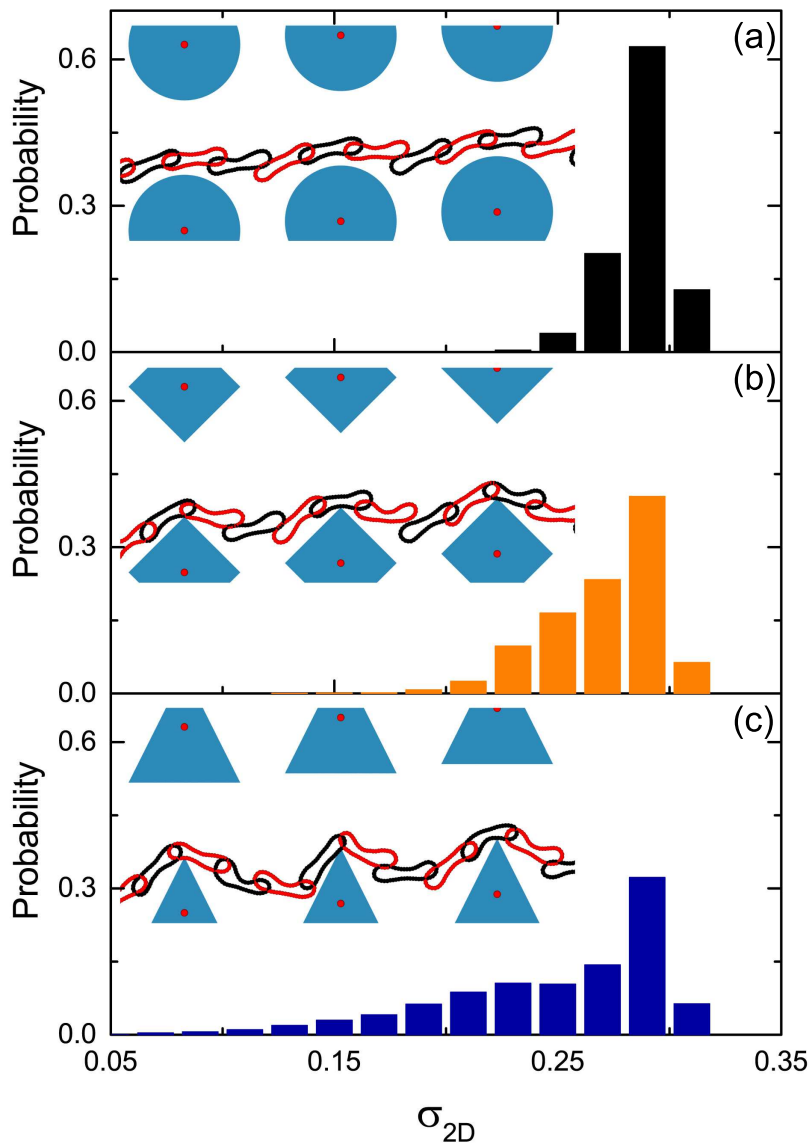}
\caption{{\bf RBC deformation in DLDs with different post geometries.} Acircularity distributions for RBCs ($K^*_{\rm 2D}  = 4$) in DLD arrays 
with (a) circular, (b) diamond, and (c) triangular posts and $\Delta \lambda  / \lambda_L = 0.05$. The presented snapshots illustrate deformation of 
RBCs in the displacement mode, see also Movie S1. Regular alteration of black and red cells corresponds to sequential positions of RBCs with a fixed spacing in the flow direction 
($\Delta x=6.0\upmu m$). $Ca = 34.1/ K^*_{\rm 2D}=8.53$.}
\label{fig5}
\end{figure}

To verify these propositions, we first consider acircularity distributions for a RBC with $K^*_{\rm 2D}  = 4$ in DLD arrays with various post geometries, 
as shown in Fig.~\ref{fig5}. For the fixed bending rigidity, a DLD array with triangular posts induces the strongest cell deformations, which is evidenced by the 
widest acircularity distribution. All three DLD devices here have $\Delta \lambda / \lambda_L = 0.05$, such that a RBC remains in the displacement mode. This 
means that the distributions in Fig.~\ref{fig5} represent RBC deformation around different pillar structures and its subsequent shape relaxation, confirming 
that sharp-edged post shapes significantly enhance RBC deformation. A stronger cell bending around triangular posts is also directly visible from 
cell snapshots in Fig.~\ref{fig5} when compared to diamond and circular structures (see also Movie S1). 

\begin{figure}[!t]
\centering
\includegraphics[width=\linewidth]{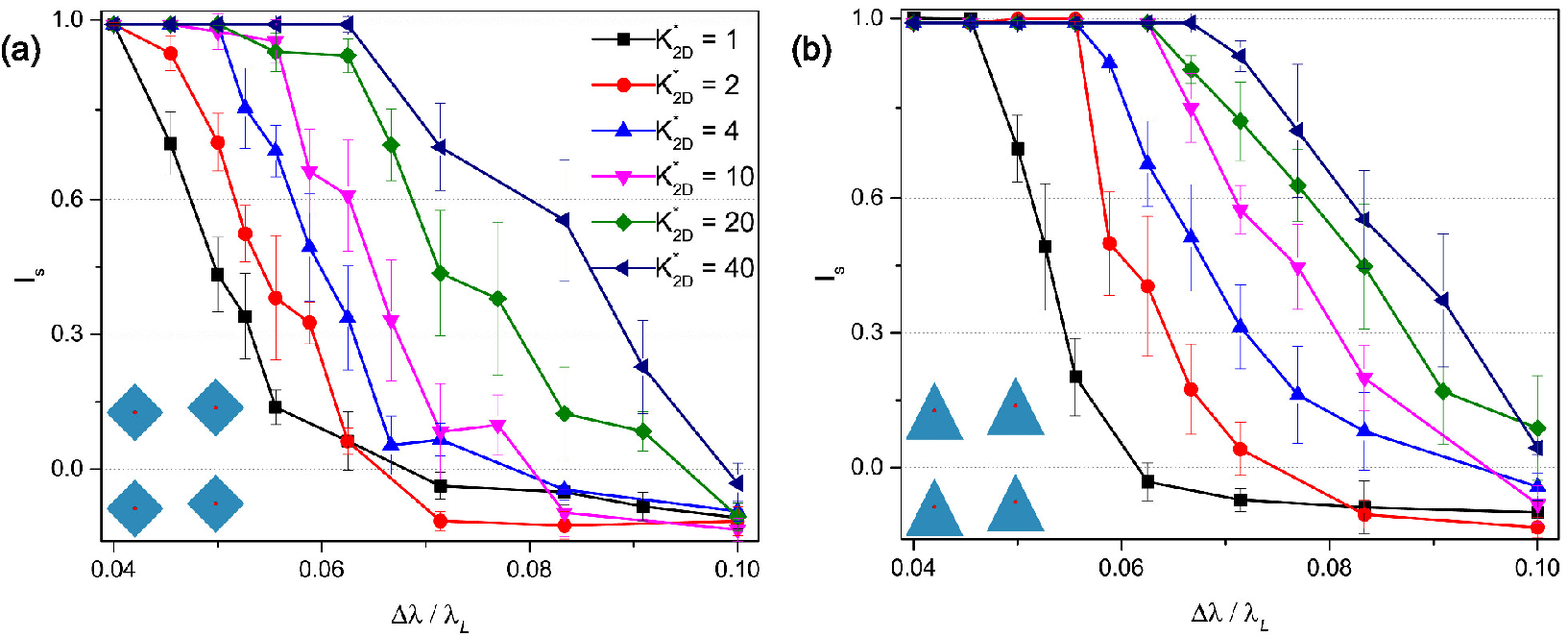}
\caption{{\bf Separation index $I_s$ of RBCs in DLDs with sharp-edged geometries.} $I_{\rm s}$ for RBCs with different bending rigidities in (a) diamond 
and (b) triangular pillar arrays as a function of the row shift $\Delta \lambda$. Error bars correspond to standard deviations. $Ca = 34.1/ K^*_{\rm 2D}$.}
\label{fig6}
\end{figure}

Figure \ref{fig6} presents separation index $I_{\rm s}$ for RBCs in diamond and triangular pillar arrays as a function of cell's rigidity and the row shift 
$\Delta \lambda$. In contrast to a very slight dependence of the displacement-to-zigzag transition on RBC bending rigidity in circular post arrays (Fig.~\ref{fig2}(b)),
this transition is much more sensitive to $\kappa$ in diamond-post arrays and most distinct for the triangular post geometry. Consequently, the diamond and 
triangular structures result in a clear improvement of the sensitivity and selectivity of a DLD device to subtle differences in RBC bending rigidity. Therefore, 
sharp-edged structures would likely lead to a practically realizable and efficient DLD design for deformability-based sorting. 

\begin{figure}[!t]
\centering
\includegraphics[width=0.6\linewidth]{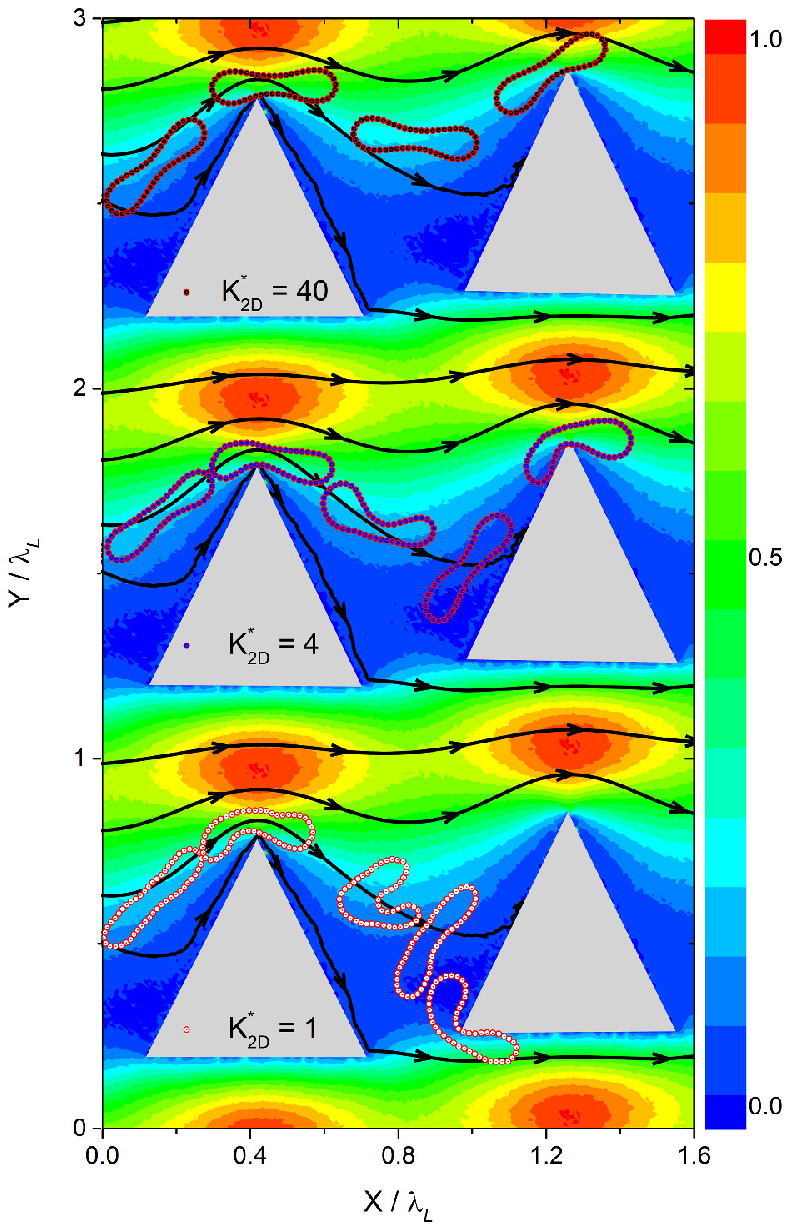}
\caption{{\bf Snapshots of RBC deformation in triangular pillar arrays.}  
Several RBC configurations for different bending rigidities extracted from flow trajectories in a DLD device with
triangular posts and  $\Delta \lambda  / \lambda_L = 0.0625$, see also Movie S2. The background shows corresponding fluid flow 
profiles of velocities in the flow direction ($x$ axis) with several typical streamlines. 
All velocities are normalized by a maximum value. $Ca = 34.1/ K^*_{\rm 2D}$.}
\label{fig7}
\end{figure}

Another interesting difference in RBC behavior in DLD arrays with circular and sharp-edged posts is that the displacement-to-zigzag transition shifts to 
a larger value of $\Delta \lambda$ with increasing $\kappa$ for circular posts, while an opposite trend is observed in arrays with diamond and triangular pillars. 
This qualitative difference indicates that there are distinct mechanisms at play. As discussed above for the circular-post array, RBC deformation 
near the top of a pillar aids in cell escaping from the first stream illustrated in Fig.~\ref{fig3} and leads to an effective increase in the critical size. Furthermore, RBC snapshots 
in Fig.~\ref{fig5} demonstrate that a RBC may not necessarily make a tight contact with circular posts, as a narrow fluid gap between the cell and the posts 
is clearly visible. This effect likely occurs due to hydrodynamic interactions of a RBC with pillar walls, which is often referred to as a 'lift' force \cite{Abkarian_TTU_2002,Messlinger_DRH_2009},
or possibly due to hydrodynamic slowing down of particles near a surface.
The lift force generally acts stronger on soft particles in comparison to rigid particles, and drives a deformable RBC away from a pillar wall, effectively increasing 
its critical size. In contrast, RBCs in diamond and triangular post arrays often make a tight contact with the posts, as seen in Fig.~\ref{fig5}, suggesting that hydrodynamic 
interactions between RBCs and the posts are weaker than in a circular pillar array. 

For a closer look at a typical RBC deformation around a triangular post, 
Fig.~\ref{fig7} shows representative cell morphologies for three bending rigidities ($K^*_{\rm 2D}=1$, $4$, and $40$) in a DLD with $\Delta \lambda / \lambda_L = 0.0625$ (see also Movie S2). 
The RBC snapshots clearly show that a softer RBC experiences a stronger bending deformation around the triangular pillar in comparison 
to stiffer cells. In fact, this deformation aids a RBC to remain within the fist stream in a DLD device and leads to an effective reduction of the critical size. 
Here, the length of the first stream in the flow direction (illustrated in Fig.~\ref{fig4}) with respect to the RBC size plays a decisive role. Thus, the stiff 
cell with $K^*_{\rm 2D}=40$, which does not bend much around the triangular post, is forced to leave the first stream due to steric RBC-edge 
interactions, since its elongated shape exceeds the length of the first stream. In contrast, the soft RBC with $K^*_{\rm 2D}=1$ is able to bend 
sufficiently around the triangular post, in order to remain within the first stream and proceed with a zigzagging mode. Therefore, the degree of RBC bending 
around a sharp edge determines whether the cell proceeds with a displacement or zigzag mode, resulting in different traversal paths in a DLD depending on 
$\kappa$. This induced deformation around a sharp edge serves as an effective sensor for RBC deformability in the DLD device.  

\begin{figure}[!t]
\centering
\includegraphics[width=\linewidth]{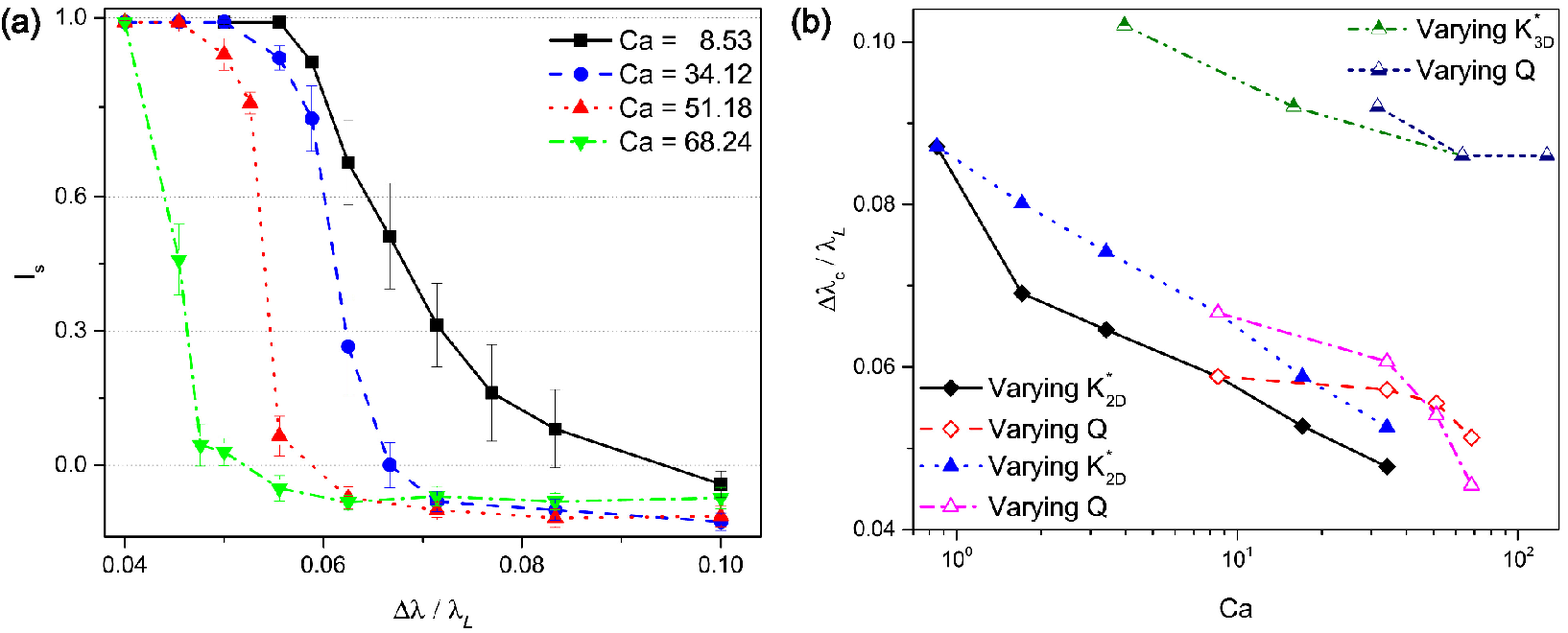}
\caption{{\bf Dependence of RBC sorting on flow rate and bending rigidity.} (a) Separation index $I_{s}$ of RBCs ($K^*_{\rm 2D}  = 4$) in triangular post arrays 
for different $Ca$ numbers (flow rate is varied) as a function of the row shift $\Delta \lambda$. Error bars correspond to standard deviations. (b) Critical row shift $\Delta \lambda_c$ (at $I_s = 0.5$) for RBC separation as a function of capillary number $Ca$. The diamond and triangle 
symbols represent diamond and triangular post arrays, respectively.  }
\label{fig8}
\end{figure}

\subsection{Effect of flow rate}

Besides RBC deformability, the flow rate in a DLD device is another important parameter, as it is directly related to applied fluid stresses. 
Intuitively, an increase in the flow rate should have a similar effect as a decrease in cell's bending rigidity for a fixed flow rate, as both lead to increased deformation. 
Figure \ref{fig8}(a) shows the separation index $I_{\rm s}$ for a RBC with $K^*_{\rm 2D}  = 4$ at various flow rates in triangular post 
arrays. Indeed, an increase in the flow rate results in a gradual shift of the displacement-to-zigzag transition to lower row shifts, 
analogously to a decrease in cell rigidity. The $I_{\rm s}$ curves in Fig.~\ref{fig8}(a) also indicate 
that the correspondence between the flow rate (or fluid stresses) and cell's rigidity needs to be selected with care. For instance, 
the differences in $I_{\rm s}$ curves for $Ca=8.53$ and $34.12$ are not as large as for $Ca=34.12$ and $68.24$, indicating that the flow rates
for $Ca \lesssim 34$ are too small to induce significant deformation for a RBC with $K^*_{\rm 2D}  = 4$, which is a key point for deformability-based sorting. Similarly, too high
flow rate would likely induce very strong deformations independently of the RBC rigidity, compromising sorting sensitivity. Therefore, 
the flow rate in a DLD device has to be carefully selected to induce distinguishable deformation for RBCs within a desired range 
of membrane rigidities.

To examine the discussed similarity of effects of cell's bending rigidity and the flow rate in DLD device in more detail, we plot
in Fig.~\ref{fig8}(b) the critical shift $\Delta \lambda_c$ (at $I_s = 0.5$) for RBC separation as a function of capillary number $Ca$.
The comparison is performed between the curves ``varying $K^*_{\rm 2D}$'' and the corresponding curves ``varying $Q$''
for same post shapes. Even though the critical shift $\Delta \lambda_c$ decreases with increasing Ca number for both ``varying $K^*_{\rm 2D}$''
and ``varying $Q$'' cases, the correspondence between these curves is only qualitative. This means that capillary number, although an important
parameter, is not the only dimensionless number which determines deformability-based separation. There are several possible reasons
that may lead to this complex behavior of RBCs, which cannot be fully characterized by a single Ca number. These include non-linear
cell deformation, non-linear relaxation of the cell shape, and complex flow field and distribution of local stresses. As a result,
both cell's rigidity and the flow rate in DLD need to be carefully considered, if a quantitative prediction of $\Delta \lambda_c$
is aimed for. 

Note that the curves in Fig.~\ref{fig8}(b) allow the selection of a proper flow rate $Q$ for a pre-defined $\Delta \lambda$
to achieve good sorting efficiency of RBCs whose bending rigidity is close to a targeted $\kappa$ value. Similarly, we can determine
a range of cell rigidities which can be targeted well for pre-selected $\Delta \lambda$ and $Q$.             

\begin{figure}[!t]
\centering
\includegraphics[width=\linewidth]{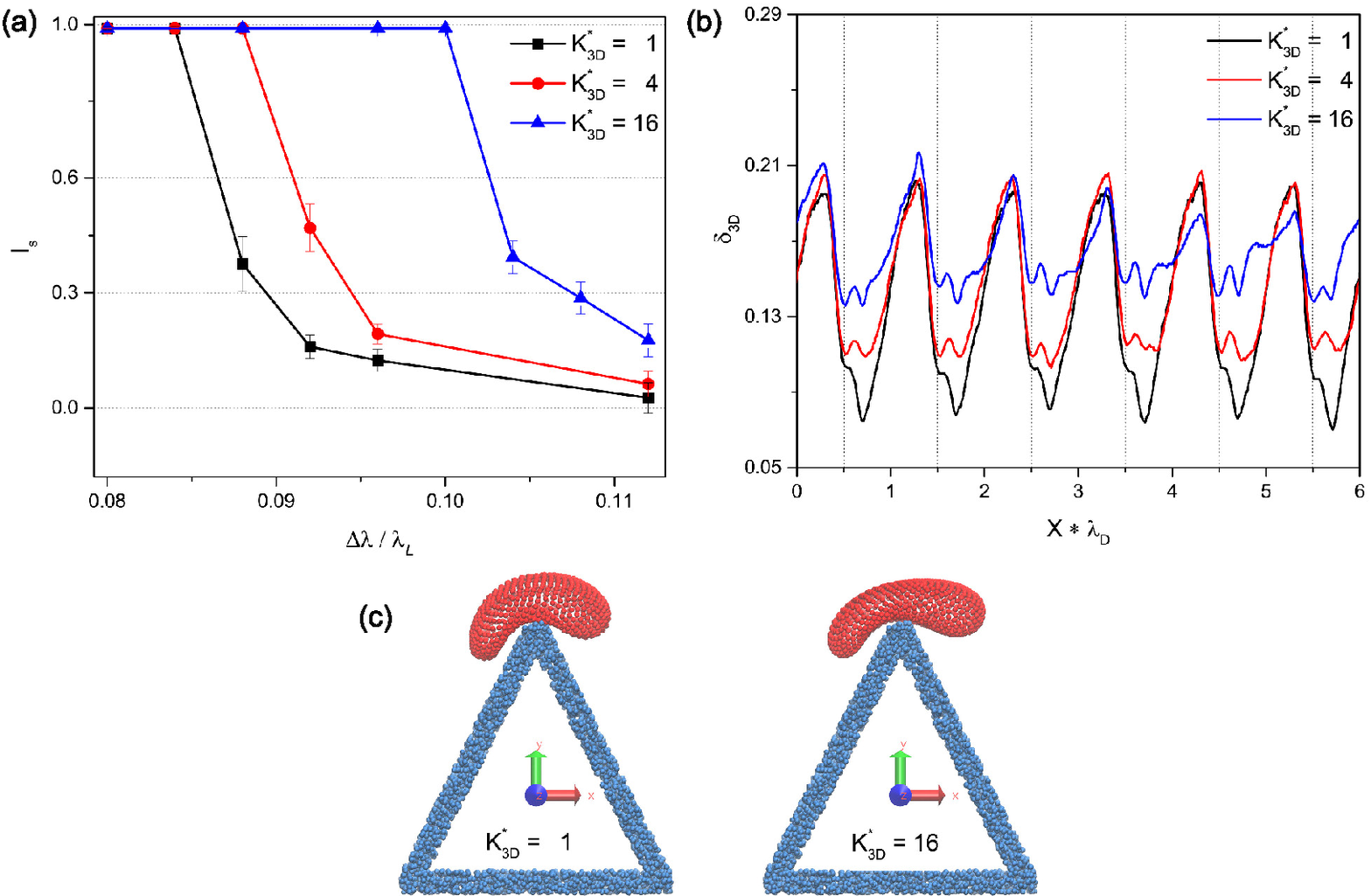}
\caption{{\bf Separation properties of RBCs with different bending rigidities in 3D.} (a) Separation index $I_{s}$ for RBCs ($K^*_{\rm 3D} = 1$, $4$, and $16$) 
in triangular post arrays as a function of the row shift $\Delta \lambda$. Error bars correspond to standard deviations. (b) Typical evolutions of the instantaneous asphericity for 
RBCs moving in the displacement mode in a device with $\Delta \lambda / \lambda_L = 0.08$. 
(c) Two representative snapshots for $K^*_{\rm 3D} = 1$ and $16$, illustrating cell bending around the top edge of triangular pillars in a device 
with $\Delta \lambda / \lambda_L = 0.08$, see also Movies S3 and S4. $Ca = 63.4/ K^*_{\rm 3D}$.}
\label{fig9}
\end{figure}

\subsection{Triangular post arrays in 3D simulations}

To verify the discussed effect of sharp-edged structures, simulations of DLDs with triangular posts have been performed in three spatial dimensions. The post geometry and arrangement of DLD are same 
as in the 2D study, while the height of the device was set to $H_{DLD}=11$ $\upmu\rm m$, such that a $8$ $\upmu\rm m$ RBC has full orientational freedom. The post edges were slightly 
blunted by replacing the corners with short planes of $1$ $\upmu\rm m$ length. This shape modification improves the stability of simulations and mimics better pillar structures 
in microfluidic devices, since very sharp corners cannot easily be achieved due to fabrication limitations and/or subsequent surface functionalization. As in the 2D simulations, bending 
rigidity of RBCs is varied such that $K^*_{\rm 3D} = \kappa/ \kappa_0 \in \{ 1...16\}$, where $\kappa_0 = 70 k_{\rm B} T$ represents a typical bending rigidity of 
a healthy RBC. The viscosity contrast between RBC cytosol and suspending medium is set to unity, in order to reduce effects from dynamic shape changes at high viscosity contrasts 
\cite{Henry_SCP_2016,Lanotte_RCM_2016} and focus primarily on cell deformation around a sharp edge. The flow rate is characterized by a Capillary number $Ca = 63.4/ K^*_{\rm 3D}$.

Figure \ref{fig9}(a) shows the separation index for RBCs with different bending rigidities as a function of the row shift $\Delta \lambda$. A comparison with Fig.~\ref{fig6}(b) demonstrates qualitative similarity between
2D and 3D results. Difference in the displacement-to-zigzag transition for $K^*_{\rm 3D} = 1$ and $4$ is quite small ($\sim 100$ nm), which is due to a relatively 
high flow rate such that RBCs strongly deform independently of its bending rigidity. This proposition is corroborated by the instantaneous asphericity in Fig.~\ref{fig9}(b), which shows 
only moderate differences in cell asphericity between $K^*_{\rm 3D} = 1$ and $4$. However, the difference in the displacement-to-zigzag transition for  
$K^*_{\rm 3D} = 4$ and $16$ is large enough ($\sim 400$ nm) to achieve deformability-based sorting in practice. Consistently, Fig.~\ref{fig9}(b) shows that 
RBC asphericities for  $K^*_{\rm 3D} = 4$ and $16$ are quite distinct, indicating an appropriate correspondence between RBC rigidity and applied fluid stresses. 
Figure \ref{fig9}(c) illustrates RBC snapshots for $K^*_{\rm 3D} = 1$ and $16$, where a noticeable difference in cell bending around the top edge of a triangular 
post is clearly visible (see also Movies S3 and S4). Finally, critical row shifts $\Delta \lambda_c$ for varying bending rigidity and flow rate
are plotted in Fig.~\ref{fig8}(b) and show a good qualitative correspondence, consistent with the 2D results. In 3D, due to the presence of shear elasticity
in RBC membrane, the second capillary number $Ca_{\mu}$ based on the shear modulus (see Section \ref{sec:sim}) may contribute to the complex cell behavior in DLDs.
In conclusion, the 3D simulation results confirm the relevance of the same separation mechanisms 
discussed in the context of $2D$ systems and support encouraging perspectives for efficient deformability-based cell sorting in DLDs with sharp-edged obstacles. 

\section{Conclusions}

Our simulation study demonstrates advantages of sharp-edged post structures in deep DLD devices for deformability-based cell sorting. It is clear that not any deformation can be 
efficiently exploited for cell separation in DLDs. Only deformations near the posts, which affect the cell positioning with respect to the separatrix between displacement and zigzag 
modes, are important. This point is best illustrated by RBC acircularity in circular post arrays in Fig.~\ref{fig2}(a), which is strongly affected by RBC bending rigidity, 
but the majority of deformation is due to flow-field inhomogeneity away from the posts, so that deformability-based sorting remains poor. Thus, cell deformation near the top of a pillar
(region of highest shear rate) governs cell's traversal through a DLD device, since it can facilitate or prevent a zigzagging event. Interestingly, this type of deformation is rather insensitive 
to RBC bending rigidity in circular post arrays. In contrast, DLDs with diamond- and triangular-shaped posts induce a favorable mode of deformation in the critical flow region, 
so that sharp-edged geometries serve as efficient deformability sensors.

The sorting capabilities of DLD devices are generally characterized by a critical size, which is related to the thickness of the first stream and defines the displacement-to-zigzag 
transition \cite{Huang_DLD_2004,Inglis_CPS_2006}. This single parameter functions perfectly for rigid spherical particles in DLDs with various geometric structures \cite{Zhang_BRD_2015}. 
Obviously, the behavior of anisotropic and deformable particles in DLDs is much more complex and cannot be described well by a single predefined critical size \cite{Zeming_RSP_2013,Henry_SCP_2016}, 
since particle deformation may alter its effective size and affect the particle trajectory. In addition to the effective particle size in comparison to thickness of the first stream, 
our simulations show that the relation between effective particle length and the length of the first stream in the flow direction is very important. For example, a stiff RBC 
in triangular post arrays is not able to bend much around a sharp edge (see Fig.~\ref{fig7}), and thus keeps its full length during cell-post interaction. However, this steric interaction
forces the stiff RBC to leave the first stream, because its length is smaller than the RBC size. In contrast, soft RBCs, which can bend enough around a sharp edge, effectively 
reduce their length and remain within the first stream, as illustrated in Fig.~\ref{fig7}. This constitutes a new mechanism that is different from the concept of a critical size and 
can be at play as well for long enough rod-like anisotropic particles. 

Cell deformation in a DLD device is directly associated with applied fluid stresses controlled by the device geometry and flow rate. Low flow rates, which do not induce substantial 
cell deformation, are not very sensitive to deformability differences. Another strong drawback of low flow rates is a low device throughput, which needs to be high 
enough for practical applications in order to process a required amount of sample within an acceptable time limit. However, too high flow rates can also be problematic, because 
even though they are advantageous for 
device's throughput, the sensitivity of deformability measurements might be compromised, as large fluid stresses can induce very strong cell deformations independent of cell rigidity. 
Thus, the flow rate (or fluid stress) has to be carefully selected, such that a proper deformation is induced to efficiently differentiate cell rigidities 
within a range of interest. Furthermore, for a decision about the device throughput, a potential trade-off between device's throughput and sensitivity needs to be considered. 

Another practical limitation of high flow rates and consequently strong cell deformations is potential cell lysis. An average shear stress in our 3D simulations can be 
estimated as $\bar{\dot{\gamma}}\eta = \kappa Ca /  D_{r}^3$, whose value is smaller than $0.1$ Pa. In fact, RBCs are able to sustain much larger stresses without lysis 
in microvasculature \cite{Popel_MH_2005} or in microfluidics \cite{Beech_SCS_2012}. A further concern is whether sharp edges might impose very strong local stresses on RBC membranes. 
The analysis of simulated RBC deformation within DLD devices with triangular posts shows that RBCs experience the strongest membrane stresses near sharp edges.
Interestingly, the largest local stresses are at the top side of a RBC that is opposite to the side of direct contact between the cell and the post. However, 
the magnitude of these stresses is maximum 40\% larger than the average fluid stresses on a RBC within a DLD device. In addition, the fabrication of very sharp edges 
is not possible in practice, so that they are slightly blunted, which leads to a  reduction of local shear stresses. Therefore, we conclude that RBC lysis is 
unlikely to be a serious issue for flow rates, which represent a good sensitivity for RBC deformability measurements.

In summary, sharp-edged structures in deep DLD devices can be employed as deformability sensors, as they induce a bending-like deformation around a sharp corner. 
Two different mechanisms contribute to this process. The first corresponds to a relation between the effective cell size and thickness of the first stream, which is the 
conventional mechanism for the characterization of DLD devices with a critical particle size. The second mechanism is related to the correspondence between effective 
cell length and the length of the first stream in the flow direction, which is altered during cell bending around a sharp edge. Optimal sensitivity for deformability 
measurements requires fine tuning of the flow rate in order to induce an appropriate deformation strength. This means that generally there exists a compromise 
between device throughput and sensitivity for deformability-based sorting. DLD devices with sharp-edged structures are expected to be applicable in situations 
where cell's mechanical properties are altered, for instance, in diseases such as diabetes, malaria, and sickle-cell. We hope that our proposition will be tested 
experimentally in the near future.

\section*{Acknowledgments}
We would like to thank Kushagr Punyani, Stefan H. Holm, Jason P. Beech, and Jonas O. Tegenfeldt (Lund University, Sweden) for stimulating discussions. 
We acknowledge the FP7-PEOPLE-2013-ITN LAPASO -- "Label-free particle sorting" for financial support. Dmitry A. Fedosov acknowledges funding by 
the Alexander von Humboldt Foundation. The authors gratefully acknowledge the computing time granted through JARA-HPC on the supercomputer JURECA 
at Forschungszentrum J\"ulich.


%

\end{document}